**A polarizing situation: Taking an in-plane perspective for next-generation near-field studies**

P. James Schuck[1,‡], Wei Bao[1,2], Nicholas Borys[1]

[1]Molecular Foundry, Lawrence Berkeley National Lab, Berkeley, CA, USA
[2]Department of Materials Science and Engineering, U. C. Berkeley, Berkeley, CA, USA
Corresponding author. E-mail: [‡]pjschuck@lbl.gov

By enabling the probing of light–matter interactions at the functionally relevant length scales of most materials, near-field optical imaging and spectroscopy accesses information that is unobtainable with other methods. The advent of apertureless techniques, which exploit the ultralocalized and enhanced near-fields created by sharp metallic tips or plasmonic nanoparticles, has resulted in rapid adoption of near-field approaches for studying novel materials and phenomena, with spatial resolution approaching sub-molecular levels. However, these approaches are generally limited by the dominant out-of-plane polarization response of apertureless tips, restricting the exploration and discovery of many material properties. This has led to recent design and fabrication breakthroughs in near-field tips engineered specifically for enhancing in-plane interactions with near-field light components. This mini-review provides a perspective on recent progress and emerging directions aimed at utilizing and controlling in-plane optical polarization, highlighting key application spaces where in-plane near-field tip responses have enabled recent advancements in the understanding and development of new nanostructured materials and devices.

# 1 Introduction

Near-field optical imaging and spectroscopy have undergone a renaissance of sorts, paralleling the past decade's enormous progress in nanoscale material design, structuring, synthesis, and control. The primary reasons for this are two-fold: (i) nano-optical approaches provide the opportunity to probe and control light–matter interactions and properties at the functionally relevant length scales of nanostructured materials; and (ii) the continued development of near-field probes has benefited greatly from recent advances in plasmonics, nanofabrication, and nanoassembly. Indeed, continually expanding near-field capabilities, combined with greatly reduced technical limitations, offer more access to physical and chemical information that is unobtainable with other methods.

At the core of all near-field studies is the near-field tip (also known as a near-field probe), which acts as a transducer for far- to near-field light and vice versa, ideally providing an approximate optical "delta function" response (relative to a far-field diffraction-limited spot) at its apex [1-13]. More specifically, the tip modifies the local density of optical states within the zeptoliter mode volume surrounding its apex [1, 14-16]. Because it is responsible for mediating all nano-optical interactions with the sample, the quality and characteristics of all near-field images critically depend on the tip [17]. It can function either as a source of localized (near-field) light or as a reporter of the sample's near-fields, and frequently acts as both. Either way, it is important to note that, upon optical excitation, the confined near fields surrounding both the sample and tip are highly complex. The fields have both electric (E) and magnetic (H) components consisting of a three-dimensional (3D) complex-valued polarization state (vector field) [18-21]. Notably, although near-fields are evanescent in nature, the concept of their degree of polarization can be well defined [22]. As a consequence, the tip's interaction with a sample (and its ability to probe and influence local sample properties) is strongly dependent on its response to different electromagnetic field polarizations. *Thus, understanding and engineering the polarization response of near-field probes is crucial for unlocking the full potential of near-field microscopy, and ultimately for investigating and developing new materials and their properties.*

Indeed, knowledge and control of the probe's polarization behavior is important for a wide range of applications. Besides the more-conventional polarimetry near-field scanning optical microscopy (NSOM) [23, 24], these include: full vector-field mapping of the nano-light surrounding next-generation nanophotonic devices [17]; mapping and controlling the radiative/nonradiative relaxation kinetics, dipole orientations and directional emission from nanoscale emitters [7, 25-38]; nano-crystallography investigations [39, 40]; and probing local nonlinear optical behavior [41-51] and magneto-optical effects [52-69]. Studies involving nanoscale vibrational spectroscopy and imaging are perhaps where the tip polarization response is most important, since a well-understood near-field polarization can reveal unique nanoscale properties such as molecular symmetry, intermolecular interactions, and bond orientations within a sample [40, 70-74]. This information is critical for developing technologies based on bio- and nanomaterials, where, for example, a tiny difference in molecular orientation can be of great functional importance [75]. In all cases, it is important to understand the polarization properties of the near-field tip for correctly interpreting the images or spectra that are recorded.

The original nano-aperture-based tip design enabled production of and interaction with near-fields that were polarized predominantly in the plane of the sample surface (in-plane polarization) [**Fig. 1(A)**] [1, 9, 76-79]. However, the major signal-to-noise constraints and spatial resolution limitations of these tips (typically restricted to ~100 nm or larger) lead to *apertureless* NSOM approaches, in which the nanoscale aperture is replaced by (a much smaller) metallic particle [80] or a sharp metal-coated scan-probe tip [1]. Notably, the polarizability of these apertureless tips is principally in the out-of-plane or "z" direction [**Fig. 1(B)**] along the long-axis of the tip [1, 7, 40, 81]. Therefore, the advantages in resolution and signal enhancement afforded by apertureless tips, as well as the advances in commercially available instrumentation [82-84], mean that *a majority of recent near-field optical investigations have concentrated primarily on z-polarized interactions with samples*. While the tips are beneficial for probing a number of key interfacial properties (e.g. molecular adsorption), the exploration and discovery of many material properties (e.g. phonons and excitons in two-dimensional (2D) materials) requires interaction with in-plane near-field light components. This has led to recent design and fabrication breakthroughs in near-field tips engineered specifically for enhanced, higher-resolution in-plane interactions (see **Fig. 1**).

This mini-review provides a perspective on recent developments aimed at utilizing and controlling *in-plane* optical polarization within near-field studies of nanostructured materials. The discussion will begin with a brief description of the polarization response of conventional apertureless near-field tips. We will then highlight key application spaces where in-plane near-field polarization studies have enabled recent advancements. Finally, we will discuss emerging prospects for controlling and converting between different near-field polarization states, as well as future directions directly impacted by better understanding, control, and enhancement of in-plane optical near-fields.

## 2 Polarization properties of apertureless tips

As noted above, sharp, metal near-field tips demonstrate a strongly anisotropic polarizability favoring the z-direction. This is clear for elongated or ellipsoidal tips oriented normal to the surface. However, even for a perfectly spherical metal nanoparticle at a tip apex, interactions with a sample (and broken symmetry along the axial direction) – with the tip/particle effectively coupling with an image dipole beneath the sample surface [**Fig. 1(B)**] – lead to preferential polarizability, local field enhancement, and scattering for z-polarized fields [4]. A phenomenological model has been successfully employed to describe the polarization-selective enhancement of apertureless tips, especially in the context of references [39, 85, 86]. Here, the Raman scattering tensor $R$, which describes the far-field polarization selection rules for Raman scattering, is modified by "tip-amplification tensors" $F$ and $F'$ to account for the polarization-selective enhancement of the excitation and subsequent scattering into the far-field, respectively. In the presence of a tip, the tip-enhanced Raman scattering (TERS) tensor becomes: $R_{TERS} = F^T R F'$. The values of the tip-amplification tensors depend on the exact geometry of the tip in a manner similar to its polarizability [48], and they are typically assumed to have the form:

$$F(\omega) = \begin{bmatrix} F_x & 0 & 0 \\ 0 & F_y & 0 \\ 0 & 0 & F_z \end{bmatrix}, F' = F(\omega') = \begin{bmatrix} F'_x & 0 & 0 \\ 0 & F'_y & 0 \\ 0 & 0 & F'_z \end{bmatrix}$$

where $F_z > F_x, F_y$ and $F'_z > F'_x, F'_y$. Here, $\omega$ and $\omega'$ are the incident optical frequency and scattered output signal frequency, respectively. Often, the frequency difference $\omega - \omega'$ can be small relative to the plasmon resonance width of the tip, such that $F \cong F'$.

Such tip-enhancement tensors well describe the polarization selectivity that can be achieved with an apertureless probe with the assumption that the off-diagonal elements = 0 proving adequate for many applications. Z-polarized apertureless near-fields have proven to be powerful, e.g., for Raman-based nano-crystallography studies [39], TERS measurements of compositional and strain gradients [42, 74, 87-104], nano-IR spectroscopic imaging of proteins and other soft materials [105-111], and fluorescence mapping of oriented molecules [25, 28, 31-33, 112-117]. The sensitivity is further enhanced by using second harmonic generation and other nonlinear optical contrast mechanisms for probing broken symmetries in materials (e.g. ferromagnetic domains) at the nanoscale [44, 46, 47, 67, 118]. Importantly, in nearly all these cases, contributions from the more weakly enhanced *x,y-polarization in:x,y-polarization out* configuration are neglected.

However, many critical electronic and vibrational excitations – particularly in lower-dimensional nanomaterials – are sensitive only to in-plane field polarizations. Additionally, detailed light–matter interactions within the tip's near-field can be considerably more complex than those described by the above tensors. For example, in **Figs. 1(A-C)**, note that the polarization is "pure" (either completely in-plane or out-of-plane) only directly along the tip axis [**Fig. 1(B)**] or in the center of the aperture or gap [**Fig. 1(A,C)**], whereas all other locations experience depolarization (i.e. the off-diagonal elements in $F$, $F'$ are not 0). Further polarization mixing results from a tip's finite cone or pyramid angles, tip tilt, and nanoscale tip inhomogeneities [86, 119, 120]. Phenomenological models can capture the cross-polarization effects of cone/pyramid angles and tip tilt quite well, and it has even been shown that a higher signal contrast in TERS experiments is possible under cross-polarized detection (z polarized excitation, in-plane detection) [86, 119].

## 3 Measuring near-field tip polarizability

Nearly all near-field tips exhibit some degree of structural heterogeneity at the nanoscale, adding variability to their near-field properties. Thus, while theoretical simulations – both analytical and numerical – can qualitatively describe the polarization response of model near-field tips, the desire for a more quantitative understanding of this has led to the development of techniques for performing near-field polarization analysis on real probes. This is particularly relevant for some recent tip designs, where a degree of intentional heterogeneity, or "controlled roughness", is exploited for improved light capture and enhancement properties in TERS measurements [**Fig. 1(F)**] [71].

From the 1990s, researchers were able to use single molecules as probes to determine the orientation of near fields [e.g. **Figs. 1(D,E)**] [26, 121, 122]. However, despite providing extremely high-resolution spatial information, single-molecule near-field fluorescence measurements are not particularly high throughput ones; hence other tip-characterization approaches were developed. For example, in 2007, Lee *et al.* showed that a conventional ellipsometry method, the so-called rotational analyzer ellipsometry technique, was capable of determining the polarizability tensor of Au-nanoparticle-functionalized tips [**Fig. 1(G)**] [123]. In this method, the tensor is built up serially by systematically rotating the polarizer angle in the scattered-light detection path for all incident light polarizations. More recently, Mino et al. showed that a defocused imaging approach could be used for establishing the tip apex polarizability based on a single measured scattering pattern [**Fig. 1(F)**] [71]. Others have employed back focal plane imaging to measure the dipole orientation of metal nanoparticles [124] and other structures [125-130]. Compared to defocused imaging, back focal plane imaging is advantageous when the tip polarization/dipole is primarily in-plane. However, defocused imaging is better for probing z-oriented dipoles and is less sensitive to laser speckle noise from Raleigh scattered light. In all cases, it was established that once the tip polarizability is determined, it becomes possible to *quantitatively* interpret near-field TERS [71] and vector-field maps [19, 123] obtained with the same tips – a key goal of near-field microscopy.

## 4 Probing in-plane modes of nanomaterials

As mentioned above, some classes of polarization-sensitive measurements such as single-molecule fluorescence and magneto-optical studies are amenable to the use of conventional aperture-based NSOM probes and thus, their in-plane polarization response. However, many other spectroscopy methods, particularly chemical imaging via vibrational spectroscopy, require signal enhancement that is achievable only with apertureless and other more-advanced tips.

*4.1 Nano-mapping of in-plane vibrations in carbon nanotubes and graphene*

Carbon nanotubes (CNTs), along with being considered a fundamental "nano building block" for many potential technologies [131], are a prototypical nanomaterial whose spectroscopic properties require probing with in-plane polarization components [**Figs. 2(A-F)**] [71, 74, 132]. In particular, the

G+ Raman band at ~1580 cm$^{-1}$ [**Figs. 2(B,C)**], a longitudinal optical phonon related to C-C bond stretching, is most sensitive to polarization along the tube axis. The Raman spectra of this and other related G modes can indicate external stresses acting on a CNT and provide information on CNT chirality. Meanwhile, the radial breathing mode (RBM) [**Fig. 2(B,D)**], with energies in the ~100–400 cm$^{-1}$ range, responds primarily to z-polarized field components and provides a direct measure of the CNT diameter. Additional Raman modes include the disorder-induced D band at ~1300 cm$^{-1}$, as well as the in-plane-sensitive Z-breathing modes [133] along the CNT axis with low and intermediate energies corresponding to the lengths of short tubes or tube segments [**Fig. 2(B)**] [134-141]. In fact, with such well-defined modes, CNTs are excellent test beds for understanding in- vs. out-of-plane near-field tip polarization properties [71].

Near-field capabilities have proven invaluable for CNT characterization, enabling numerous nanoscale insights into local CNT physics, structure, and behavior [**Figs. 2(A,E,F)**]. For example, TERS and nano-photoluminescence (nano-PL) measurements have revealed local distortions of the CNT lattice by a negatively charged defect [142], as well as local changes in chirality and conductivity [143]. However, the weaker in-plane polarizability and field enhancement capabilities of conventional apertureless probes, combined with larger background signals, have limited the extension of these types of near-field optical characterization techniques to graphene [102, 144-147] (though nano-IR measurements have proven invaluable for probing plasmons in graphene and CNTs [148-154]). The phonon properties of graphitic films are dominated by the planar symmetry of the material, making the G phonons relatively inaccessible to the conventional TERS response [145]. Still, broken symmetries, defects, and edges can result in stronger out-of-plane TERS coupling and hence, have recently resulted in some exciting investigations of nanoscale graphene properties [**Fig. 2(G)**] [145, 146].

The clear advantages of in-plane polarized near-fields for CNT (and graphene) TERS studies have motivated the use of more sophisticated probes [5, 6, 155, 156] based on optical nanoantennas [157-160]. Such probes include bowtie-antenna tips [e.g. **Fig. 1(J) and Fig. 2(A)**] [6, 30, 161, 162] and resonant coaxial antenna tips [e.g. **Fig. 1(K)**] [84, 163]. In CNT studies, these tips demonstrated both significant in-plane TERS signal enhancement while using *dielectric* substrates (previous enhancements of similar magnitude were achieved only with metal tips over metal substrates in the so-called tip-substrate gap mode) [6] and the ability to probe the Z-breathing intermediate-frequency Raman modes, which are typically very weak or absent in conventional Raman spectroscopy of CNTs [133].

*4.2 Hyperspectral nano-imaging of exciton properties in two-dimensional transition metal dichalcogenides*

Due to their remarkable light absorption and emission properties – and the unparalleled opportunities for dynamically controlling them – 2D monolayer transition metal dichalcogenides (ML-TMDCs) are ideal building blocks for atomically thin, flexible optoelectronic devices [164-176]. They exhibit unique properties and physics not seen in other 2D materials; for example, unlike graphene, many members of the TMDC family have appreciable direct bandgaps, typically in the ~1–3 eV range [175-178]. They are therefore genuine semiconductors that interact strongly with near-infrared and visible light (e.g., a monolayer of $MoS_2$ absorbs up to ~30% of all incident light with above-gap energies), making them particularly appealing for a plethora of light absorbing and/or emitting device applications [165, 172, 179-181]. The near-bandedge optical transitions in ML-TMDCs are dominated by excitons and trions (charged excitons) [182]. And unlike traditional 2D quantum wells, the enhanced coulombic interaction between the electrons and holes in ML-TMDCs stabilizes the excitonic states at room temperature with exceptionally large binding energies (~0.5–1 eV) [177, 178, 183, 184].

Unfortunately, the performance of the active ML-TMDC materials is often far below theoretical expectations, particularly for critical factors such as carrier mobility and quantum yield [165, 185]. This can be traced to a few key factors, with the primary one being a notable lack of nanoscale characterization studies, especially ones pertaining to optical properties; to date, nearly all optical investigations of ML-TMDCs have been diffraction-limited ones. This is related to the fact that the primary excited state absorbers/emitters in these systems – excitons – are polarized *within the plane of the 2D layer* [186], making them less sensitive to conventional apertureless tip near-field characterization approaches. This, combined with larger background signal issues inherent in 2D material studies, means near-field studies on ML-TMDCs and related materials been successfully realized only recently [187-190].

These investigations are enabled by next-generation near-field probes engineered for in-plane polarizability. The Campanile tip geometry [**Fig. 1(C)**] has proven particularly useful [191, 192], recently facilitating hyperspectral mapping of nanoscale excited state relaxation processes in $MoS_2$ [**Figs. 2(I,J)**] [193]. These nano-optical studies succeeded in determining and visualizing optoelectronic and excitonic properties, heterogeneity, and band-bending at the most relevant and important length scales in these materials. The (mesoscopic) effects of grain boundaries on these properties were directly imaged and quantified, with significant implications for device design. Most notably, near-field probing led to the discovery of a surprising new form/phase of $MoS_2$ at the edge region of all synthetic/chemical vapor deposition-grown materials. This is particularly important, as it constitutes a paradigm shift from only metallic states in the interpretation of edge-related physics and photochemistry in synthetic 2D materials, and has a profound impact on both catalytic applications and device technologies, as it is a critical consideration for establishing electrical contact [194].

These studies represent only the beginning of (in-plane polarized) near-field efforts aimed at elucidating the rich and unique nanoscale physics within these exciting 2D systems, ultimately guiding the future development of high-quality layered materials and next-generation devices that are expected to impact an incredibly broad range of applications.

## 5  Mapping nanoscale electric and magnetic vector fields

Lower-dimensional material systems, such as those described above, often act as the basic building blocks for novel nanophotonic device structures that are designed to control and transform light at deeply subwavelength scales. Owing to the needs of an ever-expanding application space that requires finer nano-optical control and expanded functionalities, it is critical to map the complex EM fields surrounding these device elements for characterizing, understanding, and engineering their nano-light properties. As highlighted in a recent review [17], great strides in this area of nano-optical characterization now allow researchers to probe different phase and amplitude vector components of both electric and magnetic fields at the nanoscale. Needless to say, the in-plane as well as out-of-plane field components should be effectively measured for obtaining a complete picture of how these nanophotonic elements interact and ultimately perform together.

Over the past 10 years, researchers have shown that standard apertureless tips can successfully map the *x*, *y* and *z* components of the electric field vector **E** (and even magnetic field intensities [195]) of a multitude of nanophotonic structures including plasmonic nanoparticles and cubes, optical antennas, split-ring resonators, and meta molecules [18-20, 123, 196-211]. Similarly, a well-characterized nanoparticle was recently used to visualize vector light-field distributions of tightly focused vector beams [212]. These experiments exploit the interference-based amplification techniques employed in scattering-type scanning near-field microscopy to measure both the field's amplitude and phase at each point in the scan [213], albeit with a much larger response to the z component, as noted above. Realizing the need for better in-plane sensitivity, Olmon *et al.* developed a probe consisting of a triangular Pt platelet oriented under a slight tilt angle at the tip apex, demonstrating good in- and out-of-plane response, as well as minimal depolarization effects [**Fig. 1(I)**] [200]. They used this probe to sensitively map the 3D vector electric near-field distribution surrounding an IR dimer optical antenna, and then take advantage of vector relationships to deduce the structure's conduction current density distribution **J** and its associated magnetic vector field **H**.

Aperture-based probes have proven to be quite adept at visualizing in-plane-polarized complex vector electric near-fields with high resolution using interferometric techniques [214, 215] to measure the amplitude and phase of **E** surrounding structures such as plasmonic nanowires, negative index metamaterials, nano-apertures, dielectric-clad waveguides and photonic crystal waveguides [216-221]. Of course, mapping the local **E** is just half the story, since novel materials (metamaterials) and nanostructures can have magnetic responses on par with electric responses, and also because local charges and currents create **E** and **H** fields that are neither orthogonal nor equal in strength, leading to a nontrivial relationship between them. Researchers realized that an aperture probe can detect the z-component of the optical magnetic field, $H_z$, near photonic structures (either by using a modified "split-ring" aperture tip [222] [**Fig. 1(H)**] or through an $H_z$-mediated interaction with localized EM fields [223-226]) and also act as a Bethe-hole analyzer capable of sensing in-plane optical magnetic fields [227-230]. This has enabled simultaneous mapping of the amplitude and phase of **E** and **H** fields, as demonstrated by le Feber *et al.*, for the near-fields of a photonic crystal waveguide [**Fig. 2(K)**] [231]. Pushing the limits of sensitivity and resolution even further, researchers have recently shown that the campanile near-field probes – with their stronger field confinement and larger field enhancement than aperture probes – can simultaneously measure in-plane **E** and out-of-plane **H** near-fields [**Figs. 2(L,M)**] [232]. Though the origins of the *H*-field responses for these various tips are still being clarified [17, 231-233], it is clear that a near-complete mapping of near-field vectors is now possible, representing a significant advance in nano-optical device characterization capabilities.

## 6 Future directions and outlook

Clearly, exciting and interacting with in-plane optical near-field polarization components is required for fully accessing nanostructured material properties at their most relevant length scales. To this end, current advances in near-field probe and nanoantenna development with significant in-plane polarizabilities and ultra-enhanced, strongly localized fields are central to a number of emerging nano-optical applications. For example, while the first calculations on optical trapping at the apex of an apertureless probe [234] have not resulted in a convincing demonstration, in-plane plasmonic aperture–antenna tips are now enabling trapping, spectroscopic interrogation, and manipulation of truly nanoscale objects [235-238] down to the size of individual proteins [239, 240]. The local gradient and scattering forces between a sharp tip's near-field and a photo-excited sample are now being exploited as a readout mechanism for a material's local chemistry and polarization [241-245]. Nonlinear vibrational nano-spectroscopies such as tip-enhanced stimulated Raman scattering [241] can now be employed to map in-plane vibrations and bond orientations (to date, only out-of-plane contributions have been probed [99]), further enhancing their chemical contrast capabilities, potentially down to the single-bond level. On-demand catalysis with molecular precision can be enabled by polarization-sensitive plasmon-enhanced hot-carrier extraction [246-249]. When integrated into a heat-assisted magnetic recording scheme [250, 251], polarization-selective tips have been proposed as potential technologies for all-optical high-density memory read-write heads exploiting nanoscale magneto-optical interactions [252, 253], Meanwhile, tips based on antennas that interact strongly with multiple *E* polarization components simultaneously [160, 254-256] would be capable of directly converting these components to circularly/elliptically polarized and even super-chiral near-fields [257-260], enabling unprecedented interactions with (chiral) nano-objects ranging from biomolecules to circular excitonic emitters within 2D TMDCs [179]. Indeed, modified next-generation probes are well-suited to act as nanophotonic-structured waveguides that efficiently collect and direct polarization-dependent emission from such excitonic photon sources, a function important for future solid-state quantum architectures [261, 262]. Of course, a number of near-field-related questions also exist, such as how can one perform near-field chemical imaging on a cell membrane or other soft material within a liquid? Can polarization at a tip apex ever be "pure" enough to sensitively measure Kerr rotations and related phenomena? Some exciting recent near-field time-resolved Kerr experiments suggest so [263], though it remains to be seen if this can be pushed from a ~100 nm resolution level to below 10 nm. Despite these and other challenges, it is evident that nearly all areas of science and technology stand to benefit from the pursuit of complete near-field polarization control.

**Acknowledgements** The authors thank our colleagues at the Molecular Foundry for stimulating discussion and assistance. Work at the Molecular Foundry was supported by the Director, Office of Science, Office of Basic Energy Sciences, Division of Materials Sciences and Engineering, of the U.S. Department of Energy under Contract No. DE-AC02-05CH11231.

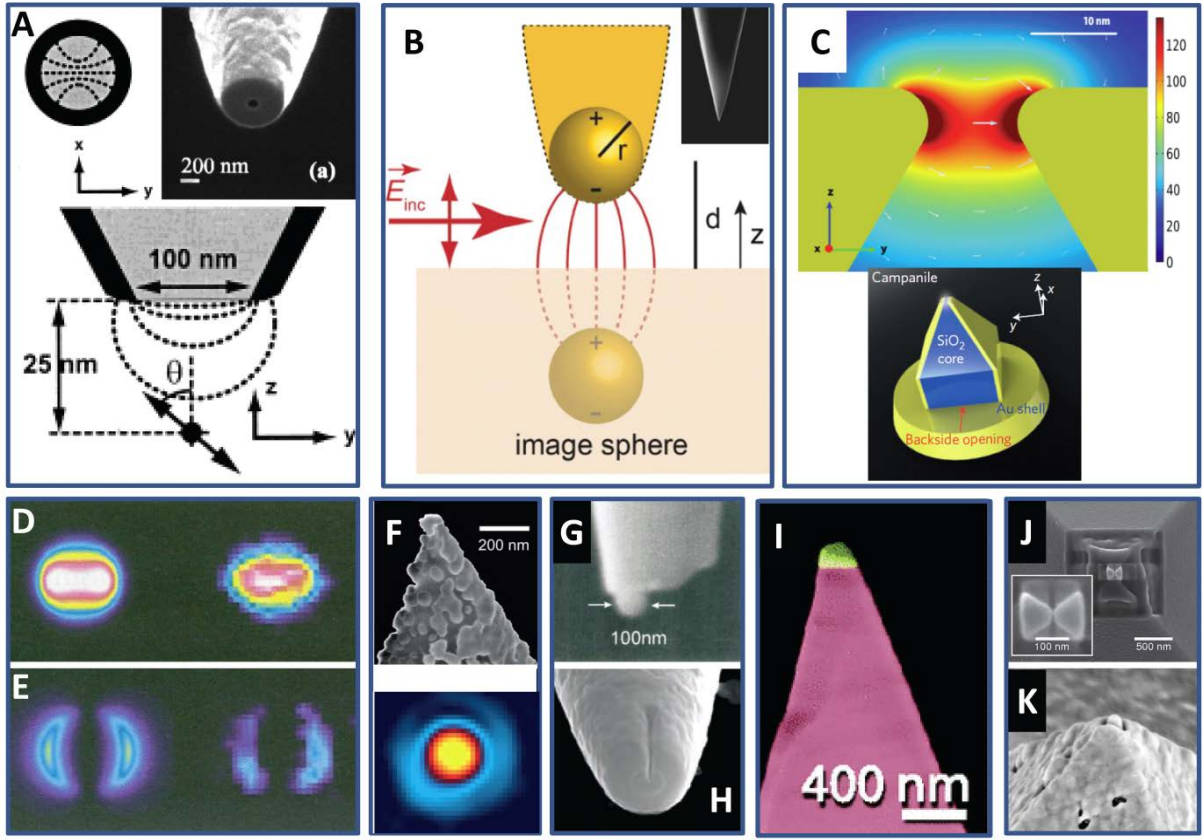

**Fig. 1 Examples of NSOM tips with different electromagnetic field polarization responses. (A)** Schematic of an aperture-based NSOM tip as seen from below (upper left) and as a section along the tip axis (bottom), as well as its associated electric field lines (dotted). Adapted from Ref. [116]. Inset: scanning electron microscope (SEM) image of an aluminum-coated aperture-based NSOM tip. Reproduced from Ref. [34]. **(B)** Model of an apertureless tip and the effective polarizability of this coupled tip–sample system approximating the tip as a sphere with radius r and tip–sample separation d and subject to an external field $E_{inc}$. Reproduced from Ref. [4]. Inset: SEM Image of a sharp metal apertureless tip courtesy of Lukas Novotny. Used with permission (www.nano-optics.org). **(C)** A *yz*-section simulation of the spatial profile of the steady state electric field amplitude near the end of a campanile tip, normalized to the incident field amplitude (top). The white arrows indicate the polarization of the electric field. A schematic of a campanile structure at the end of a gold-coated conical tapered NSOM fiber (bottom). Adapted from Ref. [191]. **(D)** Simulated $|E_x|^2$ distribution near an aperture NSOM tip (left) and the measured image of an in-plane oriented fluorescent molecule (right). **(E)** Similar to (D), but for the $|E_z|^2$ component distribution (left) and an out-of-plane oriented molecule (right); all for *x*-polarized tip excitation. (E) and (D) adapted from Ref. [26]. **(F)** SEM image of a granular metallic tip made by evaporating Ag on a silicon cantilever for AFM (top) and the experimental defocused scattering pattern from the same tip (bottom) illustrating its dipole orientation. Reproduced from Ref. [71]. **(G)** SEM image of a Au nanosphere at the end of a glass fiber tip. Reproduced from Ref. [80]. **(H)** SEM image of a split-ring aperture probe for measuring optical magnetic fields. Reproduced from [222]. **(I)** SEM image of a triangular Pt antenna probe fabricated on a Si AFM tip, with Pt highlighted in yellow and Si in red. Reproduced from [200]. **(J)** SEM image of a bowtie nanoantenna tip. Taken from [30]. **(K)** SEM image of a coaxial nanoantenna tip. Reproduced from Ref. [84].

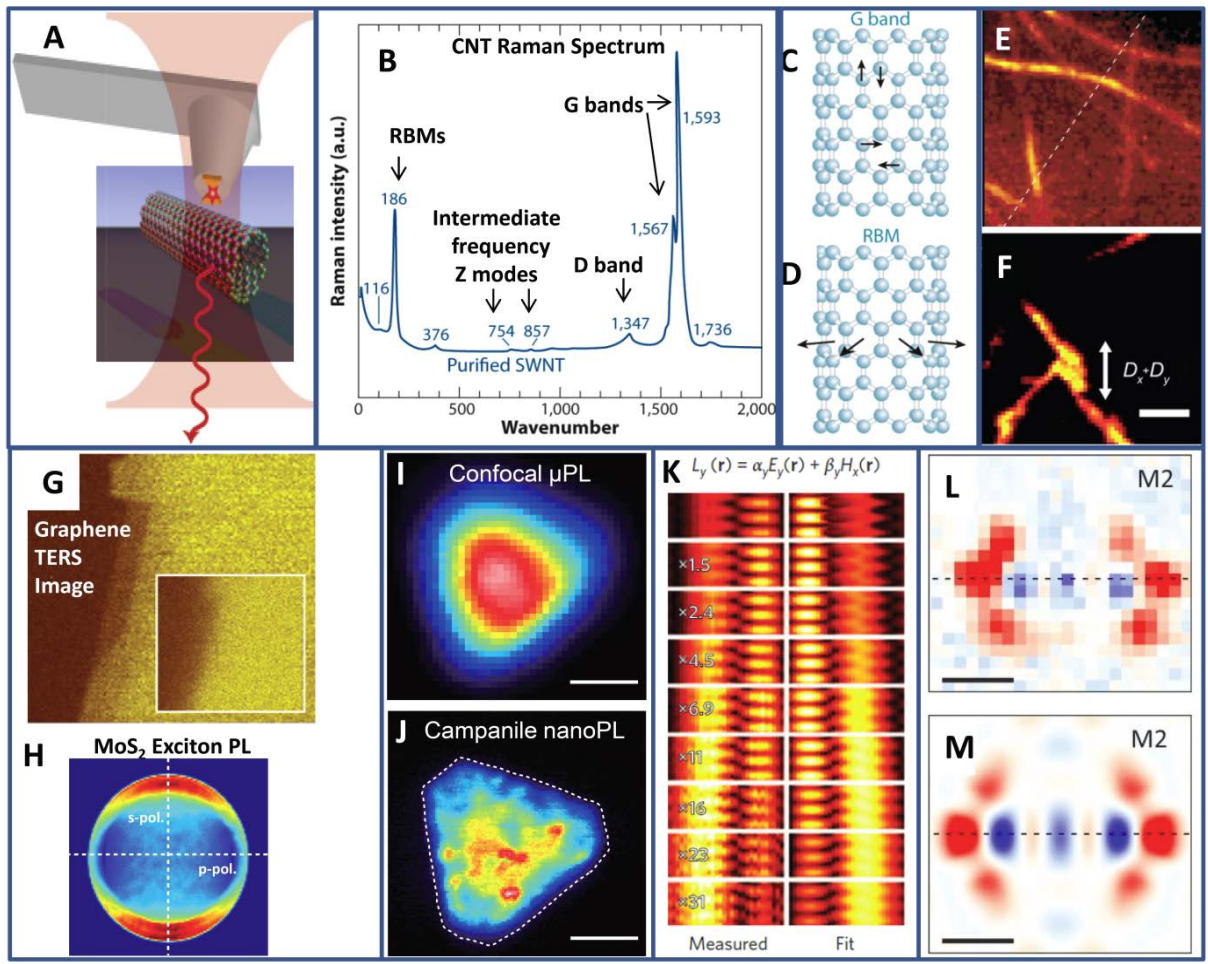

**Fig. 2 Applications of polarization-dependent NSOM.** (**A**) Schematic of a TERS measurement on a CNT using a bowtie nanoantenna tip with in-plane polarizability. Adapted from Ref. [6]. (**B**) A representative Raman spectrum from a single-wall CNT bundle. (**C**) The G-band eigenvectors for the CNT C-C bond stretching mode. (**D**) CNT radial breathing mode eigenvectors. (B-D) figures and captions adapted from Ref. [132]. (**E**) Apertureless TERS image of single-wall CNTs. Reproduced from Ref. [104]. (**F**) TERS image of single-wall CNTs taken with tip similar to that shown in Fig. 1(F). The white arrow shows the in-plane components of the tip dipole, $D_x + D_y$. Reproduced from Ref. [71]. (**G**) Graphene TERS image of the G' band. Inset: confocal image of the same area. Adapted from Ref. [146]. (**H**) Back focal plane emission pattern from a $MoS_2$ monolayer showing purely in-plane exciton dipole emission. Adapted from Ref. [186]. (**I**) Confocal micro-PL image and (**J**) campanile near-field nano-PL image of the same monolayer MoS2 flake. Adapted from Ref. [193]. (**K**) Aperture-based NSOM measurements, and fits, at different heights above a photonic crystal waveguide containing information of both in-plane optical electric and magnetic fields. Adapted from Refs. [17] and [231]. (**L**) Measured and (**M**) calculated electric (red) and magnetic (blue) field intensity distributions above a photonic crystal nanocavity. The measured fields were collected with a campanile tip. Reproduced from Ref. [232].

**References**

1. L. Novotny and B. Hecht, Principles of Nano-Optics, Cambridge: Cambridge University Press, 2006